\documentclass[twocolumn]{svjour3}          % twocolumn
\smartqed  % flush right qed marks, e.g. at end of proof
\usepackage{graphicx}

\usepackage[dvipsnames]{xcolor}
\usepackage{listings}
\usepackage{url}
\usepackage{hyperref}
\usepackage[misc]{ifsym}

\newcommand\YAMLcolonstyle{\color{red}\mdseries}
\newcommand\YAMLkeystyle{\color{black}\bfseries}
\newcommand\YAMLvaluestyle{\color{blue}\mdseries}

\makeatletter

\newcommand\language@yaml{yaml}

\expandafter\expandafter\expandafter\lstdefinelanguage
\expandafter{\language@yaml}
{
  keywords={true,false,null,y,n},
  keywordstyle=\color{darkgray}\bfseries,
  basicstyle=\YAMLkeystyle,                                 % assuming a key comes first
  sensitive=false,
  comment=[l]{\#},
  morecomment=[s]{/*}{*/},
  commentstyle=\color{purple}\ttfamily,
  stringstyle=\YAMLvaluestyle\ttfamily,
  moredelim=[l][\color{orange}]{\&},
  moredelim=[l][\color{magenta}]{*},
  moredelim=**[il][\YAMLcolonstyle{:}\YAMLvaluestyle]{:},   % switch to value style at :
  morestring=[b]',
  morestring=[b]",
  literate =    {---}{{\ProcessThreeDashes}}3
                {>}{{\textcolor{red}\textgreater}}1
                {|}{{\textcolor{red}\textbar}}1
                {\ -\ }{{\mdseries\ -\ }}3,
}

\lst@AddToHook{EveryLine}{\ifx\lst@language\language@yaml\YAMLkeystyle\fi}
\makeatother

\newcommand\ProcessThreeDashes{\llap{\color{cyan}\mdseries-{-}-}}

\begin{document}

%\title{TOSCA-based orchestration of complex infrastructures in heterogeneous clouds}
\title{Orchestrating complex application architectures in heterogeneous clouds}




\author{Miguel Caballer\and
        Sahdev Zala \and
        {\'A}lvaro L{\'o}pez Garc{\'i}a \and
        Germ{\'a}n Molt{\'o} \and
        Pablo Orviz Fern{\'a}ndez \and
        Mathieu Velten}

\institute{
        A. L{\'o}pez Garc{\'i}a \and P. Orviz Fern{\'a}ndez \at
        Instituto de F{\'i}sica de Cantabria. \\
        Centro Mixto CSIC - UC. Santander, Spain.\\
        \email{aloga@ifca.unican.es}           %  \\
    \and
        Miguel Caballer \and Germ{\'a}n Molt{\'o} \at
        Instituto de Instrumentaci{\'o}n para Imagen Molecular (I3M). \\
        Centro Mixto CSIC - Universitat Polit{\`e}cnica de Val{\`e}ncia. Valencia, Spain.
    \and
        Sahdev Zala \at
        IBM Raleigh. North Carolina. USA.
    \and
        Mathieu Velten \at
        European Organization for Nuclear Research (CERN). Geneva, Switzerland.
}

\date{
This is an author's, pre-print version of this work.
The final publication is available at link.springer.com via
http://dx.doi.org/10.1007/s10723-017-9418-y
}

\maketitle

\begin{abstract}
Private cloud infrastructures are now widely deployed and adopted across
technology industries and research institutions. Although cloud computing
has emerged as a reality, it is now known that a single cloud provider
cannot fully satisfy complex user requirements. This has resulted in a
growing interest in developing hybrid cloud solutions that bind together
distinct and heterogeneous cloud infrastructures. In this paper we
describe the orchestration approach for heterogeneous clouds that has been
implemented and used within the INDIGO-DataCloud project. This
orchestration model uses existing open-source software like OpenStack
and leverages the OASIS Topology and Specification for Cloud Applications
(TOSCA) open standard as the modelling language. Our approach uses
virtual machines and Docker containers in an homogeneous and transparent
way providing consistent application deployment for the users. This approach is
illustrated by means of two different use cases in different scientific communities,
implemented using the INDIGO-DataCloud solutions.

\keywords{Cloud-Computing \and Heterogeneous-Cloud \and Multi-Cloud \and Open-Source \and TOSCA}
\end{abstract}

\section{Introduction}

The scientific exploitation of cloud resources is nowadays a reality.
Large collaborations, small groups and individual scientists have
incorporated the usage of cloud infrastructures as an additional way of
obtaining computing resources for their research. However, in spite of this large adoption,
cloud computing still presents several functionality gaps that make
difficult to deliver its full potential, specially for scientific usage \cite{Ramakrishnan2010,Ramakrishnan2011}.
One of the most prominent challenges is the lack of elasticity and transparent
interoperability and portability across different cloud technologies and
infrastructures \cite{Petcu2014,Lorido-Botran2014,Galante2016}. It is absolutely
needed to provide users with seamless dynamic elasticity over a large pool
of computing resources across multiple cloud providers.

Commercial providers can create this illusion of infinite resources
(limited by the amount of money that users can afford to pay), but this is
not true in scientific datacenters, where resources tend to be more limited
or are used in a more saturated regime \cite{Ramakrishnan2011,Stockton2017}.
In this context, it is a clear requirement that a user application must be
capable of spanning over several different and heterogeneous
infrastructures as a way to obtain the claimed flexibility and elasticity.
Orchestrating multiple IaaS (Infrastructure as a Service) resources
requires deep knowledge on the infrastructures being used, something
perceived as too low level by normal scientific users. However, it is
possible to hide this complexity moving the user interaction
up in the cloud model stack. This way users do not deal anymore with infrastructure
resources, but rather interact with PaaS (Platform as a Service) or SaaS
(Software as a Service) resources. In these cases, the orchestration
complexity is carried out by the middleware layer that provides the platform
or software as a service, thus the low level details can be
hidden to the users.

%overcoming this particular limitation the problem described here by
In this work we will describe how the INDIGO-DataCloud project \cite{web:indigo} is overcoming
this limitation by providing a mechanism to orchestrate computing resources across
heterogeneous cloud infrastructures. We will also thoroughly describe how
this solution is being exploited to deliver the execution of complex
scientific applications to the final users. INDIGO-DataCloud
is an European Union's Horizon 2020 funded project that aims at developing a data
and computing platform targeting scientific communities, deployable on multiple
hardware and provisioned over hybrid (private or public) e-infrastructures. INDIGO-DataCloud
is helping application developers,
e-infrastructures, resource providers and scientific communities to overcome current
challenges in the cloud computing, storage and network areas, being the
orchestration across heterogeneous providers one of the project's main objectives.

The remainder of this paper is
structured as follows. In Section~\ref{sec:background} we describe related work in the area.
In Section~\ref{sec:vision} we describe the INDIGO-DataCloud overall approach including the
technology choices that the project has made. In Section~\ref{sec:orch} we include a
high-level architectural description of
the orchestration technique that INDIGO-DataCloud is implementing. Section~\ref{sec:use cases}
contains some selected use cases, in order to illustrate the architecture
previously described. Finally, we present our conclusions and the future work in
Section~\ref{sec:conclusions}.

\section{Background and related work}
\label{sec:background}

The usage and promotion of open standards (being TOSCA \cite{Lipton2013} one of them) in the cloud
as a way to obtain more interoperable, distributed and open infrastructures is a
topic that has been already discussed \cite{teckelmann2011mapping,LopezGarcia2016b}.
Major actors \cite{Koski2015} agree that these principles should drive the evolution
of cloud infrastructures (specially scientific clouds \cite{LopezGarcia2013}) as the
key to success over closed infrastructures.

As a matter of fact, the European Commission recommended, back in 2004, the
usage of Open Standards in its ``European Interoperability Framework for
pan-European eGovernment Services'' \cite{Idabc2004}. In the same line, the United
States (US) National Institute of Standards and Technology (NIST) has encouraged
US national agencies to specify cloud computing standards in their public
procurement processes \cite{Bumpus2013}. Similarly, the United Kingdom Government
provided a set of equivalent principles in 2014 \cite{web:open_standards:uk}.

However, providers and users perceive that lower level (i.e. infrastructure
provision and management) standards hinder the adoption of cloud
infrastructures \cite{Campos2013}. Cloud technologies and frameworks tend to have a fast
development pace, adding new functionalities as they evolve, whereas standards'
evolution is sometimes not as fast as the underlying technologies. This has
been perceived as a negative fact limiting the potential of a given cloud
infrastructure, that sees its functionality and flexibility decreased. On top of
this, infrastructure management is also perceived too low level when moving
to more service-centric approaches that require not only the deployment and
management of services, but also all their operational concerns (like fault or
error handling, auto-scaling, etc.) \cite{liu2011cloud}.

In this service-centric context, cloud orchestration is being considered more
and more important, as it will play the role needed to perform the abstractions
needed to deploy complex service architectures for a wide range of application domains,
such as e-government, industry and science. Cloud orchestration involves the automated
arrangement, coordination and management of cloud resources (i.e. compute, storage and
network) to meet to the user's needs and requirements \cite{Caballer2015}.
Those requirements normally derive from the user demand of delivering a service
(such as a web service where there is a need to orchestrate and compose different services
together), performing a business logic or process, or executing a given scientific workflow.

Cloud orchestration within science applications has been tackled before by several
authors, in works related to specific scientific areas such as bioinformatics and biomedical
applications \cite{Krieger2017}, neuroscience \cite{Stockton2017}, phenomenology physics \cite{Campos2013},
astrophysics \cite{Sánchez-Expósito2016}, environmental sciences \cite{Fiore2013}, engineering \cite{korambath2014deploying}, high energy physics \cite{Toor2014}, etc. This has been also addressed in more
generic approaches, not bounded to a scientific discipline \cite{Zhao2015,Kacsuk2016,Distefano2014}.
However, these works tend to be, in general, too tied to a given type of application and workload, and
they need to be generalized in order to be reused outside their original communities.

Several open source orchestration tools and services exist in the market, but
most of them come with the limitation of only supporting their own Cloud Management Platforms (CMPs)
as they are developed within those project ecosystems. As an example we can cite some of them:
OpenStack Heat \cite{web:heat} and its YAML-based Domain
 Specific Language (DSL) called Heat Orchestration Template (HOT) \cite{web:hot}, native to OpenStack \cite{web:openstack}.
OpenNebula \cite{web:opennebula} also provides its own JSON-based multi-tier cloud application
 orchestration called OneFlow \cite{web:oneflow}.
Eucalyptus \cite{web:eucalyptus} supports orchestration via its implementation of the AWS
 CloudFormation \cite{web:cloudformation} web service.
All of them are focused on the their own CMPs and furthermore
 they rely on their own DSL languages (open or proprietary ones such as CloudFormation).

Moving from the CMP specific tools and focusing on other orchestration stacks we can find:
Cloudify \cite{web:cloudify}, which provides TOSCA-based orchestration across
 different Clouds, but is not currently able to deploy on OpenNebula sites, one of the main CMPs used
 within science clouds being supported by the project.
Apache ARIA  \cite{web:aria} is a very recent project,
 not mature enough and without support for OpenNebula.
Project CELAR \cite{web:celar} used an old TOSCA XML version using SlipStream \cite{web:slipstream} as the orchestration layer (this project has no activity in the last years and SlipStream has the
 limitation of being open-core, thus not supporting commercial providers in the open-source version).
CompatibleOne \cite{yangui2014compatibleone} provided orchestration capabilities based on the Open Cloud
 Computing Interface (OCCI) \cite{Nyren2010,Metsch2010,Metsch2011}. However the project has not been
 active in the last years.
OpenTOSCA \cite{web:opentosca} currently only supports
 OpenStack and EC2 providers.

In contrast, the Infrastructure Manager (IM) \cite{web:im} supports TOSCA-based deployments over a variety of
 cloud backends including OpenNebula and OpenStack, the two main CMPs targeted in the project; commercial
 cloud providers such as Microsoft Azure \cite{web:azure}, Amazon Web Services (AWS) \cite{web:aws},
 Google Cloud Platform (GCP) \cite{web:gcp} and Open Telekom Cloud (OTC) \cite{web:otc};
 and the EGI Federated Cloud \cite{web:egi-fedcloud} a large-scale pan-european federated IaaS Cloud to
 support scientific research.

As we can see, CMP-agnostic tools tend to move away from specific DSLs and to utilize open standards
such as OCCI or TOSCA. Although both may seem suitable for orchestration purposes, OCCI is an standard
focused on all kind of management tasks \cite{LopezGarcia2016b}, whereas TOSCA is a standard designed specifically to
model cloud-based application architectures. Choosing TOSCA as the description language for an orchestration tool
is a reasonable choice, as we will later describe in Section~\ref{sec:tosca}.

\section{INDIGO-DataCloud vision}
\label{sec:vision}

The project's design specification \cite{INDIGO-DataCloud2015} has put the focus not only
on evolving available open-source cloud components, but also on developing new solutions to
cope with the project targets, introducing as a result innovative advancements
at the layer of IaaS, e.g. by implementing advanced scheduling strategies based
on fair share or preemptible instances \cite{Garcia2017}, at the layer of PaaS, e.g. by creating
SLA-based orchestration components that support deployments on multi-Clouds \cite{salomoni2016indigo}
and, finally, at the layer of SaaS, e.g. by developing high-level REST and
graphical user interfaces to facilitate the usage of computing infrastructures
for different scientific communities \cite{Plociennik2016}.

The heterogeneity in the IaaS platforms has been addressed by the adoption of
the two leading open-source CMPs, \emph{OpenStack} \cite{web:openstack}
and \emph{OpenNebula} \cite{web:opennebula}. \emph{OpenStack} is a major open-source collaboration
that develops a cloud operating system that controls large pools of compute,
storage, and networking resources throughout a datacenter. \emph{OpenNebula}
provides a simple but feature-rich and flexible solution for the comprehensive
management of virtualized data centers to enable private, public and hybrid
IaaS clouds. Both are enterprise-ready solution that include the functionality
needed to provide an on-premises (private) cloud, and to offer public
cloud services.

\begin{figure}[t!]
\centering
\includegraphics[width=1.0\linewidth]{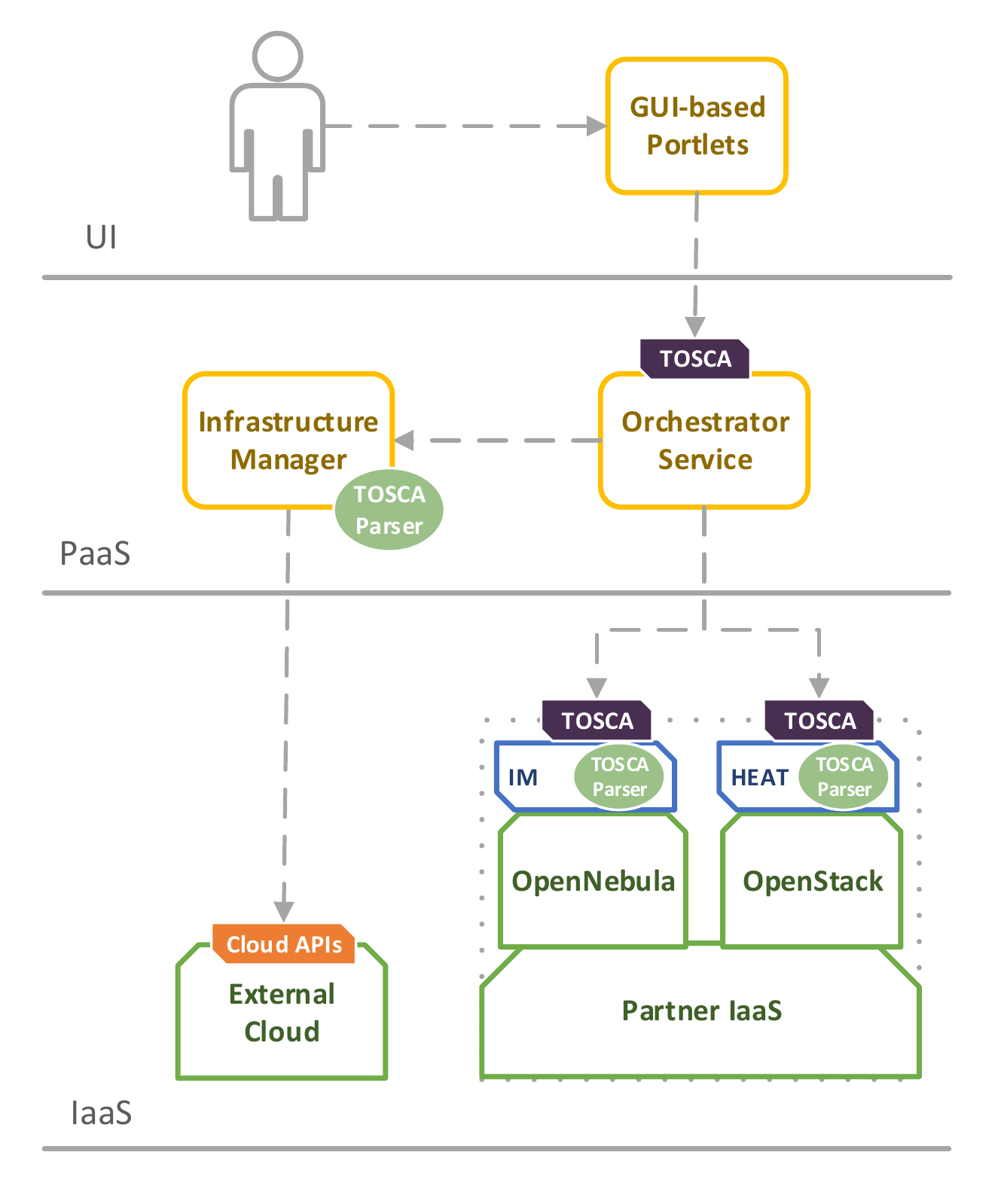}
\caption{\label{fig:indigo-arch} Simplified high level INDIGO-DataCloud architecture.}
\end{figure}

Figure~\ref{fig:indigo-arch} shows a high level and simplified overview of the
INDIGO-DataCloud architecture. As it can be seen, both CMPs at the IaaS layer
are used indistinctly from the PaaS, leveraging the TOSCA open standard at both
layers, as we will describe later in Section~\ref{sec:orch}. The different
software components supporting the TOSCA standard are key in the
INDIGO-DataCloud architecture, and the disparity in its maturity level have
brought along multiple developments in the codebase of both CMPs, in many cases
resulting in upstream contributions
\cite{web:stackalytics-tosca-parser,web:stackalytics-heat-translator}.

In the rest of this section we will elaborate on the reasons that led us to
chose the TOSCA standard, and the different existing open source components
that are being adopted and enhanced in order to support the project's use
cases.

\subsection{TOSCA}
\label{sec:tosca}

The interoperability required to orchestrate resources either in
\emph{OpenStack} or \emph{OpenNebula} from the PaaS layer has been
provided by the usage of the TOSCA (Topology and Specification for Cloud
Applications) \cite{Lipton2013} open standard. TOSCA is a
Domain Specific Language (DSL) to describe cloud application architectures, developed
by the OASIS \cite{web:oasis}
nonprofit consortium and supported by several companies as contributors,
reviewers, implementers or users. These companies include, AT\&T, Bank of
America, Brocade, Cisco, Fujitsu, GigaSpaces, Huawei, IBM, Intel, Red Hat, SAP,
VMWare, Vnomic and ZTE corporation. The usage of TOSCA was already made
practical by OpenStack projects like TOSCA Parser \cite{web:toscaparser} and Heat Translator \cite{web:heattranslator}. Both projects are easy to consume via Python Package Index (PyPI) \cite{web:pypi} packages
or directly from the master branch of the source code.

INDIGO-DataCloud adopted the TOSCA language as it allows to define interoperable descriptions of cloud
applications, services, platforms, data and infrastructure along with their
requirements, capabilities, relationship and policies. TOSCA enables
portability and automated management across multiple clouds regardless
of the underlying platform or infrastructure and is supported by a
large and growing number of international industry leaders.

TOSCA uses the
concept of \emph{service templates} to describe cloud application architectures as a \emph{topology template},
which is a graph of \emph{node types} (used to describe the possible building
blocks for constructing a service
template) and \emph{relationship types} (used to define lifecycle operations to
implement the behavior an orchestration engine can invoke when instantiating a
service template).

Three additional open software components are
taking part on the orchestration solution in the INDIGO-DataCloud project:
\emph{TOSCA Parser} is an OpenStack open-source tool to parse documents
expressed using the TOSCA Simple Profile in YAML \cite{TOSCA-YAML}. \emph{Heat
Translator} translates non-Heat templates (e.g. TOSCA templates) to the native
OpenStack's orchestration language HOT (Heat Orchestration Template). Last but
not least, the \emph{Infrastructure Manager} (IM) \cite{Caballer2015} is
a TOSCA compliant orchestrator, which relies on the \emph{TOSCA Parser}, that
enables the deployment and configuration of the virtual infrastructures over a
large set of different on-premises CMPs (e.g. OpenNebula and OpenStack) and
public Cloud providers.

\subsection{TOSCA Parser}

\begin{figure}[ht!]
    \centering
    \includegraphics[width=0.9\linewidth]{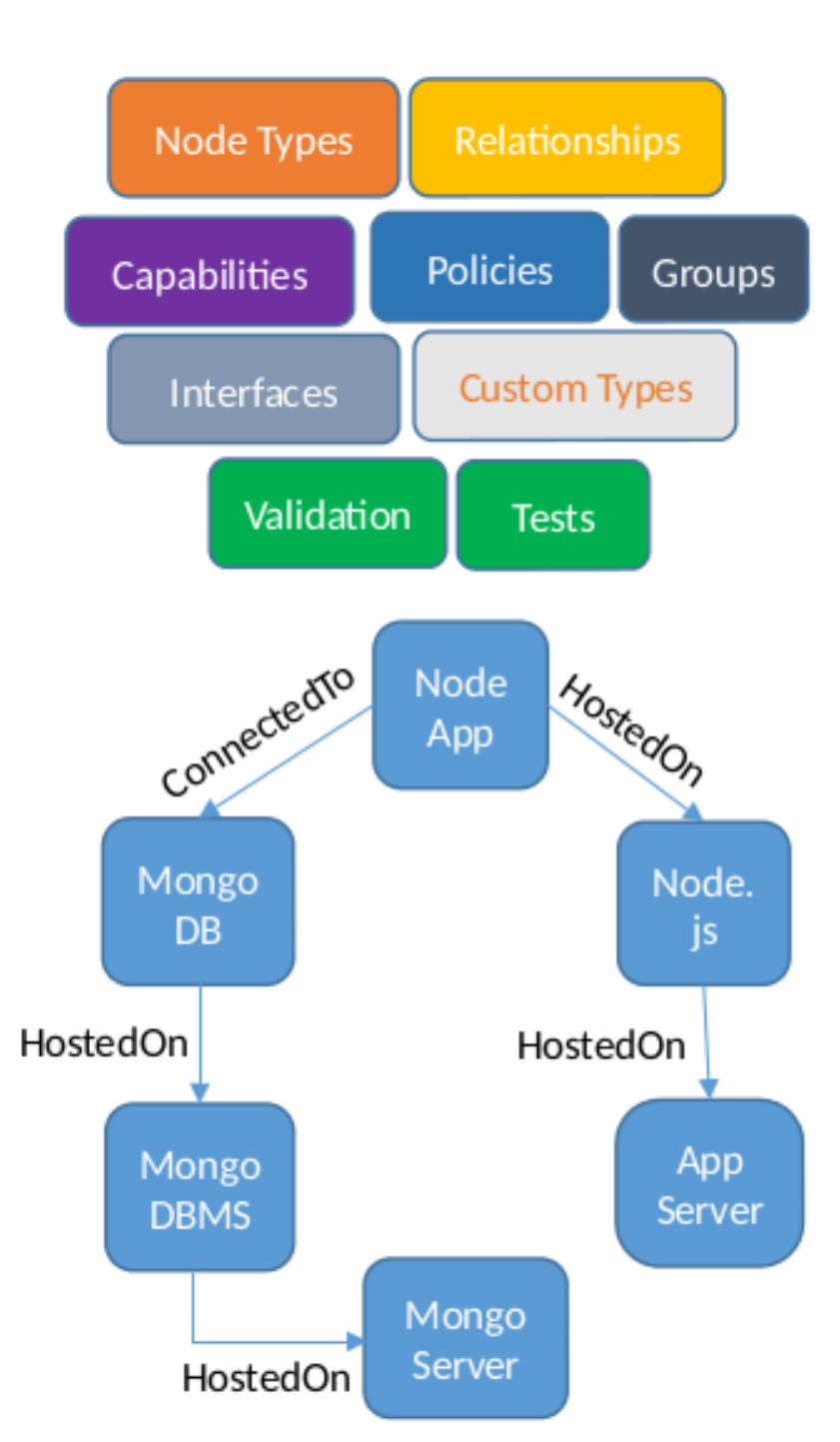}
    \caption{The in-memory representation of TOSCA nodes.}
    \label{fig:tosca-parser-nodes}
\end{figure}

The TOSCA Parser is an OpenStack project, although it is
a general purpose tool, not restricted to be used within an OpenStack
environment. The TOSCA Parser is a Python library able to read TOSCA simple
YAML templates, TOSCA Cloud Service Archive (CSAR) and TOSCA Simple Profile for
Network Functions Virtualization (NFV), creating in-memory graphs of TOSCA
nodes and their relationship, as illustrated in
Figure~\ref{fig:tosca-parser-nodes}.

\subsection{Heat Translator}

\begin{figure}[ht!]
    \centering
    \includegraphics[width=0.9\linewidth]{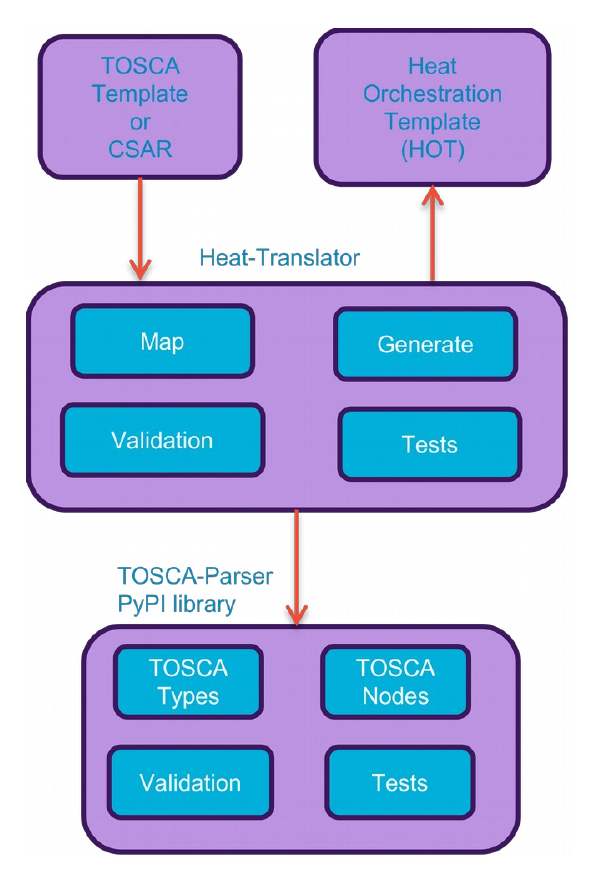}
    \caption{Heat Translator Architecture.}
    \label{fig:heat-translator-arch}
\end{figure}

The OpenStack Heat Translator project enables
integration of TOSCA into an OpenStack cloud. With Heat Translator a user can
translate TOSCA templates to the OpenStack native HOT language. These templates
can then be automatically deployed in an OpenStack cloud
(Figure~\ref{fig:heat-translator-arch}).

\begin{figure*}[ht!]
    \centering
    \includegraphics[width=0.8\linewidth]{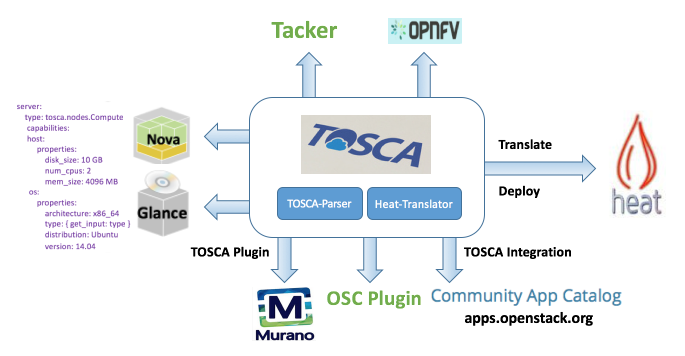}
    \caption{Heat Translator OpenStack integration.}
    \label{fig:heat-translator}
\end{figure*}

The Heat Translator project can be used directly from the OpenStack command
line and web user interface, and is well integrated into the OpenStack
ecosystem (Figure~\ref{fig:heat-translator}). It uses various OpenStack
projects for translation purposes (like image and instance type mapping) and it
is also consumed by other OpenStack official projects like the OpenStack NFV
Orchestration project, Tacker \cite{web:openstacktacker}.

Listing~\ref{Fig:TOSCAExample} shows a TOSCA document that, when passed to the
Heat Translator, results in the HOT output shown in
Listing~\ref{Fig:HOTExample}.

%\begin{minipage}{\linewidth}
\begin{lstlisting}[language=yaml, basicstyle=\footnotesize\ttfamily, breaklines=true, frame = single,
caption=TOSCA document equivalent to Figure~\ref{Fig:HOTExample}.,label=Fig:TOSCAExample]
tosca_definitions_version: tosca_simple_yaml_1_0

topology_template:
  node_templates:
    my_server:
      type: tosca.nodes.Compute
      capabilities:
        host:
          properties:
            num_cpus: 2
            disk_size: 10 GB
            mem_size: 512 MB
          os:
            properties:
              architecture: x86_64
              type: linux
              distribution: RHEL
              version: 6.5
\end{lstlisting}
%\end{minipage}

%\begin{minipage}{\linewidth}
\begin{lstlisting}[language=yaml, basicstyle=\footnotesize\ttfamily, breaklines=true, frame = single,
caption=HOT document equivalent to Listing~\ref{Fig:TOSCAExample}., label=Fig:HOTExample]
heat_template_version: 2013-05-23

parameters: {}
resources:
  my_server:
    type: OS::Nova::Server
    properties:
      flavor: m1.medium
      image: rhel-6.5-test-image
      user_data_format: SOFTWARE_CONFIG
outputs: {}
\end{lstlisting}
%\end{minipage}

\subsection{Infrastructure Manager}

%The Infrastructure Manager (IM) \cite{Caballer2015} enables to perform
%the orchestration, deployment and configuration of the virtual infrastructures
%over a large set of different cloud providers: Cloud Management Platforms such
%as OpenNebula or OpenStack, public clouds such as Amazon Web Services, Google
%Cloud Platform or Microsoft Azure and also federated clouds such as the EGI
%Federated Cloud \cite{FernandezdelCastillo2015} or FogBow
%\cite{barros2015using}.

The Infrastructure Manager (IM) \cite{Caballer2015} performs the orchestration,
deployment and configuration of the virtual infrastructures and it was chosen within
the project as it provided support for TOSCA based orchestration, as long as a
wide variety of backends and infrastructures (as described in Section~\ref{sec:background}),
specially those targeted by the project.

\begin{figure}[ht!]
    \centering
    \includegraphics[width=0.9\linewidth]{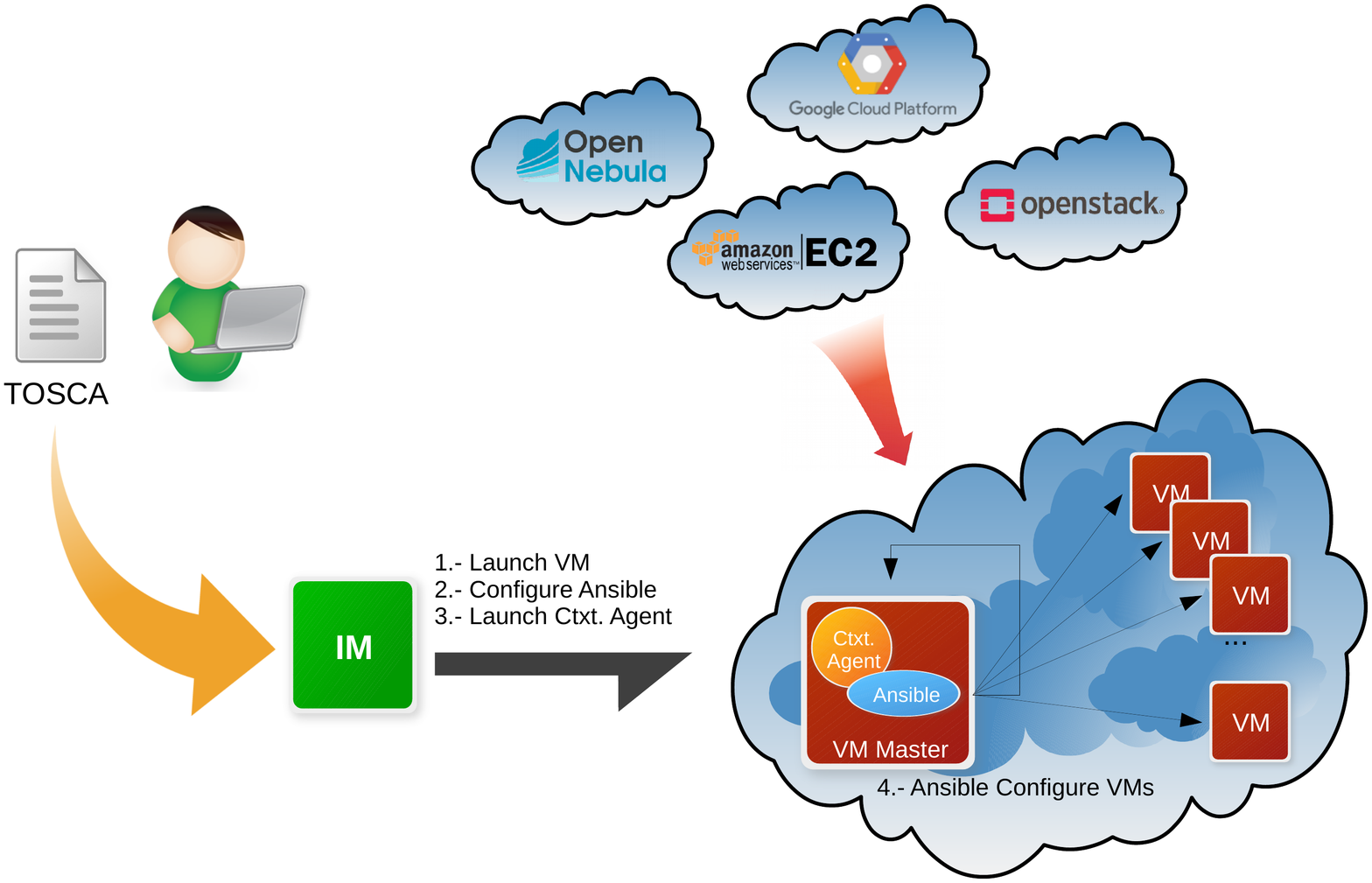}
    \caption{The Infrastructure Manager (IM).}
    \label{fig:im}
\end{figure}

Figure~\ref{fig:im} shows the scheme of the IM procedure. It receives the TOSCA
template and contacts the cloud site using their own native APIs to orchestrate
the deployment and configuration of the virtual infrastructure. The IM also
uses the TOSCA parser to parse and load in memory the TOSCA documents received
as input. Once the resources have been deployed and they are running, the IM
selects one of them as the ``master'' node and installs and configures Ansible
\cite{Ansible} to launch the contextualization agent that will configure all
the nodes of the infrastructure. The master node requires a public IP
accessible from the IM service and must be connected with the rest of nodes of
the infrastructure (either via a public or private IP). Once the node is
configured, the IM will launch the contextualization agent to configure all the
nodes using the defined Ansible playbooks. Ansible was chosen over other DevOps tools such as Puppet, Chef or SaltStack due to the combination of the following features: i) YAML support, the same language used to define the TOSCA templates, ii) Ansible Galaxy \cite{web:indigo-galaxy}, an online repository to share with the community the open-source Ansible roles created to dynamically install the services and end-user applications, iii) easy to install tool, and iv) agent-less architecture enabling the management of the nodes without requiring any pre-installed software (using standard SSH or WinRM connections).

\section{TOSCA orchestration in INDIGO-DataCloud}
\label{sec:orch}

INDIGO-DataCloud provides a comprehensive solution for deploying cloud
applications in multiple CMPs that may need complex topologies and operational
requirements, such as auto-scaling resources according to the
application needs.

As explained in Section~\ref{sec:vision}, this cloud orchestration scenario is
perfect for using TOSCA to specify resources in heterogeneous environments,
guiding the operation management throughout the application lifecycle. The tools
described in the previous section provide the functionalities needed to cope with
the orchestration needs of those cloud applications at the infrastructure and
platform level.

Basically the implemented solution enables a user to deploy cloud applications over
complex cloud infrastructures. The user interacts with a set of APIs or GUI based
portlets that enable the definition of the relevant parameters for the application
and infrastructure deployment. This is internally managed as a TOSCA document that
is sent to the different components of the architecture to manage the lifecycle of
the cloud topology: selecting the best image and cloud site to deploy the
infrastructure and then contacting the TOSCA orchestration endpoint (IM or Heat)
for the selected site.

Our approach proposes a new solution to deploy the final application at each resource
provider by combining the usage of Docker containers and VMs in a transparent way for
the user. It leverages Docker containers as the preferred underlying technology to
encapsulate user applications \cite{web:indigo-dockerhub}, but in case that containers are not supported natively
by the CMP or provider, it can also use virtual machines, achieving the same application
deployment and execution environment. This process is being handled by the orchestration
layer and is transparent for the user, resulting in the application being delivered to the user,
regardless of the using Docker containers or VMs.

Once the TOSCA document is built upon the user requirements the system starts with the
deployment of the application. In this TOSCA document, the user application is referenced,
so that the INDIGO-DataCloud Orchestrator \cite{web:indigo-orchestrator} can select the
most suitable site for executing it. After the site is selected, the orchestrator can apply
two different procedures for deploying it, based on the image availability at the selected site:

\begin{enumerate}
    \item Whenever the requested image is registered at the cloud site, the
        configuration step is removed, so the user application is spawned right
        away without the need of any image contextualization.

    %To add native support to docker containers as "first class" resources in
        %the CMPs OneDock\footnote{\url{https://github.com/indigo-dc/onedock}}
        %is used, in the case of OpenNebula, and
        %nova-docker\footnote{\url{https://github.com/indigo-dc/nova-docker}}
        %is used in the case of OpenStack.

    \item For those cases where the pre-configured image is not at the local
        catalogue of the cloud site, the deployment of the application is performed on a vanilla
        virtual machine or Docker container using Ansible roles. The execution of the Ansible Role on the provisioned computing resource is performed by either the IM, on an OpenNebula site, or Heat, on an OpenStack site. Therefore the
        application deployment is automatically done, without any user
        intervention. The Ansible role deals with the installation and
        configuration of a given application, so every application supported in
        the project need to have its corresponding role online available before
        its actual instantiation.
\end{enumerate}

A user application made available through the second approach will notably
take more time to be deployed, when compared with the pre-configured
image instantiation already described. However using Ansible
roles at this stage allows a more flexible customization since
they can be designed to support application installation on a wide set of platforms and operating system distributions. Therefore, they are not being constrained to a specific OS distribution, as
it is the case of using pre-configured images.

The availability of the Ansible role for each supported application is
taken for granted within the workflow, since the pre-configured images
are also created from them in order to install the application in a Docker image that will be made available in Docker Hub \cite{web:indigo-dockerhub}. Having a single, unified approach to
describe the application installation and configuration steps, promotes
re-usability and simplifies maintenance.

Applications being integrated in INDIGO-DataCloud require:
i) an Ansible Role that performs the automated installation of the application
together with its dependences on a specific Operating System flavour (or a subset of them);
ii) an entry in Ansible Galaxy to easily install the Ansible Role;
iii) a new TOSCA node type that defines the requirements for the application;
iv) a TOSCA template that references the new node type and optionally
specifies an existing Docker image in Docker Hub
with the application
already installed inside. This Docker image will be automatically registered in a
Cloud site supporting a Docker-enabled CMP by means of the INDIGO RepoSync tool \cite{web:indigo-reposync}. Notice that this process is
just required once. The user would later just use the same TOSCA template
to automatically provision instances of the application on-demand.

\begin{figure*}
\centering
\includegraphics[width=0.8\textwidth]{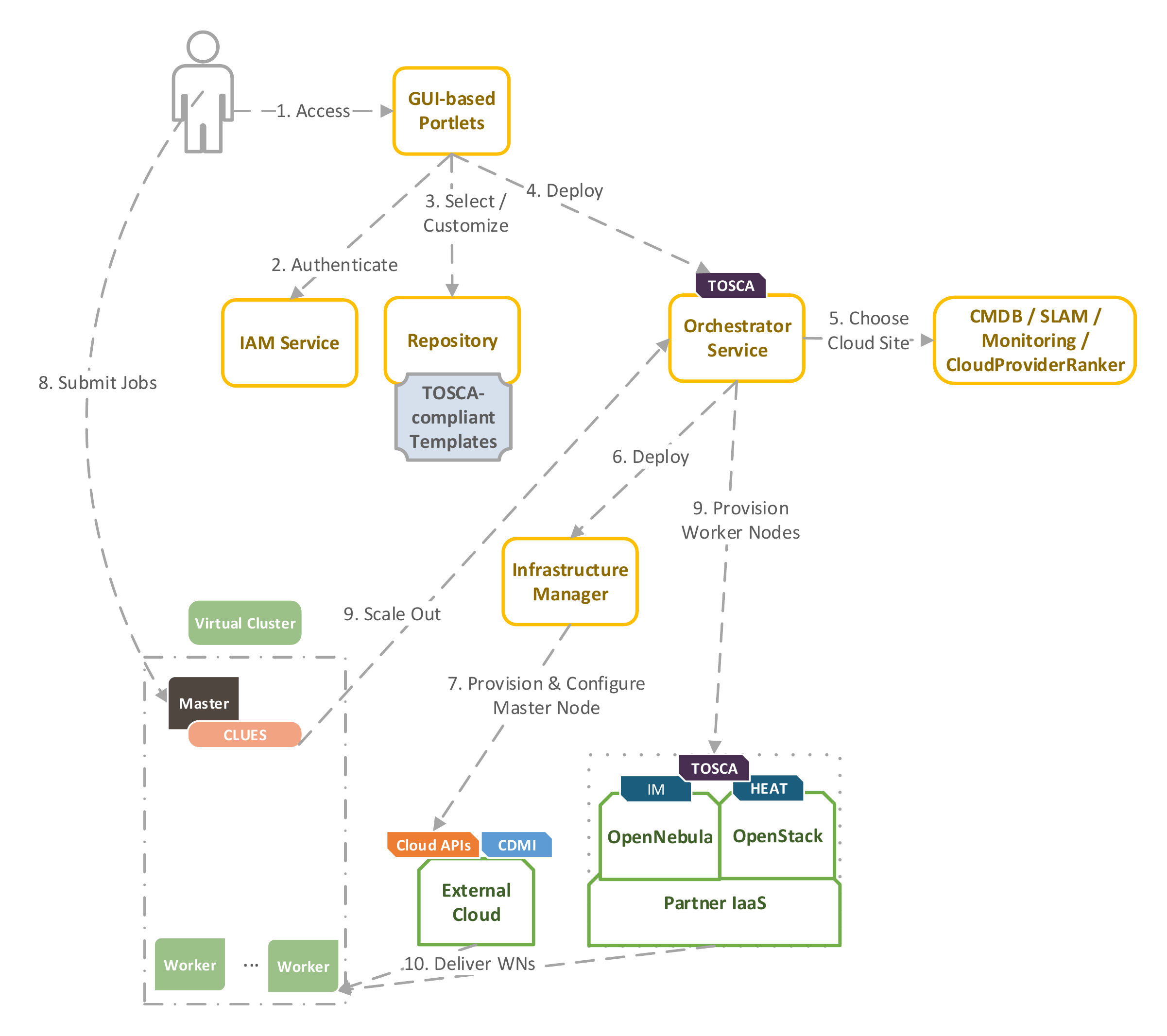}
\caption{\label{fig:tosca-arch} Simplified architecture of the usage of TOSCA for the deployment of computing clusters in INDIGO-DataCloud}
\end{figure*}

Figure \ref{fig:tosca-arch} describes a simplification of the INDIGO-DataCloud
architecture that enables the deployment of complex application layouts. In
particular, the figure describes the workflow that involves both the
TOSCA-based provision of the computing resources and their dynamic management
for the specific case of a virtual elastic cluster.

The TOSCA document is obtained out of a TOSCA template, explicitly created for
each particular application, filled with some runtime parameters. These
parameters are provided, through the
interaction with a high-level graphical user interface (GUI) (step 1). The GUI
first asks for the user credentials by means of an identity access management
service (step 2), then prompts for the selection of the TOSCA template (step
3). For the sake of example, we assume that the user wants to deploy a virtual elastic cluster.

The INDIGO-DataCloud orchestrator is the entry point to the PaaS layer,
receiving TOSCA documents as input (step 4), to find the best match for the
resource provisioning. The decision-making process is based on a SLA (Service
Level Agreement) analysis and the assessment of the potential target providers
availability (step 5), leveraging the INDIGO-DataCloud PaaS service stack \cite{salomoni2016indigo,7912692}. One important feature of the orchestration system is that it supports hybrid deployments across different public or on-premises Cloud providers, making use of the VPN technology to establish secure and seamless connections among the compute nodes located in the different infrastructures (see an example TOSCA template for a hybrid deployment in \cite{web:tosca-across-clouds}). The hybrid capabilities are particularly interesting for use cases involving virtual elastic clusters, allowing them to potentially scale out to access more servers than a single infrastructure could provide.

The CMP type of the selected cloud resource provider marks the remaining steps
in the TOSCA orchestration workflow. Whenever an OpenStack cloud provider is
selected, the orchestrator performs the interaction with the Heat service,
preceded by the TOSCA template translation by means of the Heat
Translator API. As described in Section 3, Heat Translator makes use of
TOSCA Parser utility to load the TOSCA template, ending up with a
OpenStack's native HOT description of the stack to be deployed.

On the other hand, if the target cloud provider is based on OpenNebula
framework or external public cloud platforms (such as Amazon AWS or Google
Cloud), then the Orchestrator delegates on the IM to perform the provision and the configuration of
the virtual infrastructure (step 6). The IM component will act as the
unified TOSCA orchestrator, receiving the TOSCA template and acting as the
orchestration layer with TOSCA support on top of the CMP.

The last steps in the workflow are related to the concrete example of deploying
the virtual cluster. The orchestration layer ---IM or OpenStack Heat---
performs the automated deployment of the different tools to configure the virtual
cluster along with CLUES \cite{DeAlfonso2013} as the elasticity manager of the
cluster (step 7). The user is then provided with the endpoints and credentials
to access his virtual cluster. Once the user starts submitting jobs (step 8),
CLUES automatically detects that additional resources are required, contacting
the Orchestrator (step 9). It will then restart the provisioning process of step
5, resulting in new nodes dynamically added to the existing virtual cluster
(step 10).

\begin{figure*}
\centering
\includegraphics[width=\textwidth]{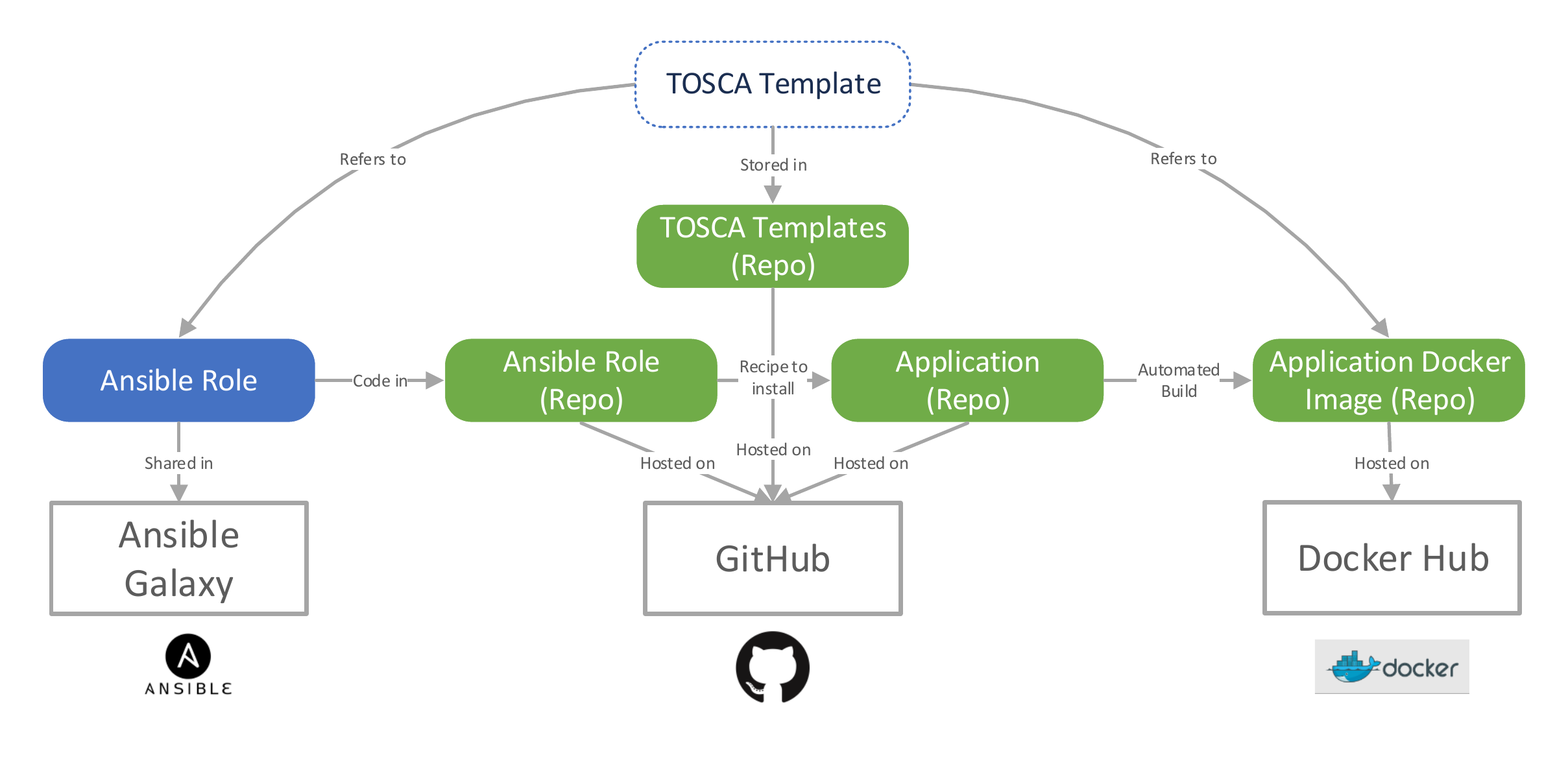}
\caption{\label{fig:tosca-repos}  Relation among Ansible Roles, Docker images and TOSCA templates.}
\end{figure*}

Figure \ref{fig:tosca-repos} describes the relation among the Ansible roles and
Docker images, as described in the TOSCA template \cite{web:indigo-tosca}. The source code of the
user applications are available in GitHub together with the Ansible roles that
describe their installation and configuration process. Profiting from the
GitHub and Docker Hub tight integration, automated builds of each application
image are triggered once a new change is committed to its repository's default
branch, thus making the last version of the application online available in
Docker Hub registry. The Ansible roles are also centrally registered in the
Ansible Galaxy online catalog. These Ansible roles are then referenced in the
corresponding TOSCA types and used in the TOSCA templates so that applications
can be automatically deployed on the provisioned virtual infrastructure.

\section{Use Cases}
\label{sec:use cases}

This section illustrates two different use cases that have been implemented
within INDIGO-DataCloud: A single node based application (Powerfit)
and an elastic Mesos cluster. Both examples use new TOSCA types added by the
INDIGO-DataCloud project to extend TOSCA Simple Profile in YAML Version 1.0.

\subsection{Powerfit}

The Powerfit \cite{powerfit} use case provides an example of a single-node
application workflow. The reduced version of the TOSCA template required to
deploy the application is shown in Listing~\ref{Fig:PowerFitTosca} (the full
example is available at {\emph{tosca-types} GitHub
repository} \cite{web:tosca-type-powerfit}).

In this use case the user selects a node of the type
\texttt{tosca.nodes.indigo.Powerfit}, defined as shown in
Listing~\ref{Fig:PowerFitDef}. The definition includes the Ansible role needed
to install and configure the application \cite{web:disvis-powerfit-role}. This is the same role used to build
the Docker image specified in the template, i.e.,
\texttt{indigodatacloudapps/powerfit}). The deployment of this application
follows the steps shown in section \ref{sec:orch}, i.e. using the pre-installed
Docker image whenever it is locally available, or otherwise using a vanilla VM
or Docker container.

%\begin{minipage}{\textwidth}
\begin{lstlisting}[language=yaml, basicstyle=\footnotesize\ttfamily, breaklines=true, frame = single,
caption=A modified excerpt of the Powerfit TOSCA template., label=Fig:PowerFitTosca]
tosca_definitions_version: tosca_simple_yaml_1_0

imports:
  - indigo_types: indigo-dc/tosca-types/master/custom_types.yaml

topology_template:

  node_templates:

    powerfit:
      type: tosca.nodes.indigo.Powerfit
      requirements:
        - host: p_server

    p_server:
      type: tosca.nodes.indigo.Compute
      capabilities:
        ...
        os:
          properties:
            type: linux
            distribution: ubuntu
            version: 14.04
            image: indigodatacloudapps/powerfit

    ...
\end{lstlisting}
%\end{minipage}

%\begin{minipage}{\textwidth}
\begin{lstlisting}[language=yaml, basicstyle=\footnotesize\ttfamily, breaklines=true, frame = single,
caption=tosca.nodes.indigo.Powerfit node type definition., label=Fig:PowerFitDef]
tosca.nodes.indigo.HaddockApp:
derived_from: tosca.nodes.SoftwareComponent
properties:
  haddock_app_name:
    type: string
    description: Haddocking application
    required: true
    constraints:
    - valid_values: [ disvis, powerfit ]
artifacts:
  galaxy_role:
    file: indigo-dc.disvis-powerfit
    type: tosca.artifacts.AnsibleGalaxy.role
interfaces:
  Standard:
    configure:
      implementation: https://raw.githubusercontent.com/indigo-dc/tosca-types/master/artifacts/haddock/haddock_install.yml
      inputs:
        haddock_app_name: {get_property:[SELF,haddock_app_name]}

tosca.nodes.indigo.Powerfit:
derived_from: tosca.nodes.indigo.HaddockApp
properties:
  haddock_app_name:
    type: string
    required: true
    default: powerfit
    constraints:
      - equal: powerfit
\end{lstlisting}
%\end{minipage}

This example demonstrates the capability to extend TOSCA with additional
non-normative types that best match the application requirements. These new
types also link to the appropriate automated deployment procedures (i.e.
Ansible roles), previously tested to guarantee the success under different
environments. Having custom types for different applications simplifies the
TOSCA templates, enhancing their readability while preventing users from
introducing changes that could affect the deterministic behaviour of the
application deployment.

\subsection{Elastic Mesos Cluster}

Using virtual Apache Mesos clusters \cite{web:apache-mesos} in the cloud enables scientific communities
to get access to customized cluster-based computing on-demand, to address both the
execution of batch jobs and long-running services via Chronos \cite{web:chronos} and the
Marathon \cite{web:marathon} framework, respectively. The project made available TOSCA templates to support the deployment of different
types of customized virtual elastic computing clusters.

These clusters are elastic since, initially, only the front-end nodes are
deployed. Moreover, unlike traditional computing clusters based on more traditional
Local Resource Management Systems (LRMS) ---like SGE, Torque or HTCondor---, Mesos provides with
a high-availability mode that features multiple masters and load balancers.

The Mesos masters are customized with the required scientific applications,
specific for a given user community. They are also configured with CLUES
support, the elasticity management system for clusters introduced in Section~\ref{sec:orch}.
CLUES monitors the state of the job queue to detect when additional Mesos
slaves are required to be deployed in order to cope with the number of pending
jobs. The cluster is then dynamically adapted to a given workload, constrained
by the maximum number of slaves specified in the TOSCA document as
\emph{``max\_instances"}. CLUES was extended within the INDIGO-DataCloud project to provision
additional nodes from the PaaS Orchestrator, as well as to introduce elasticity
for Apache Mesos Clusters and HTCondor batch resources.

An example of a TOSCA-based description for these virtual elastic computing
clusters is available in the {\textit{tosca-types} GitHub
repository} \cite{web:tosca-type-mesos},
and summarized in Listing~\ref{Fig:MesosCluster}. The TOSCA template provides a
description of the elastic cluster in terms of the computing requirements for
all the Mesos cluster nodes (master, load balancer and slave nodes). It also
specifies the maximum number of slaves to launch (5 in this example). For the sake of
simplicity, some information has been omitted in the TOSCA template depicted.

%\begin{minipage}{\textwidth}
\begin{lstlisting}[language=yaml, basicstyle=\footnotesize\ttfamily, breaklines=true, frame = single,
caption=A modified excerpt of the virtual elastic Mesos cluster TOSCA template., label=Fig:MesosCluster]
tosca_definitions_version: tosca_simple_yaml_1_0

imports:
  - indigo_types: indigo-dc/tosca-types/master/custom_types.yaml

topology_template:

  node_templates:
    elastic_cluster_front_end:
      type: tosca.nodes.indigo.ElasticCluster
      requirements:
        - lrms: mesos_master
        - wn: mesos_slave

    mesos_master:
      type: tosca.nodes.indigo.LRMS.FrontEnd.Mesos
      properties:
        marathon_password: marathon_password
        chronos_password: chronos_password
      requirements:
        - host: master_server

    mesos_slave:
      type: tosca.nodes.indigo.LRMS.WorkerNode.Mesos
      capabilities:
        wn:
          properties:
            max_instances: 5
            min_instances: 0
      properties:
        master_ips: {get_attribute:[master_server,public_address]}
      requirements:
        - host: mesos_slave_server

    mesos_load_balancer:
      type: tosca.nodes.indigo.MesosLoadBalancer
      properties:
        master_ips: {get_attribute:[master_server,public_address]}
      requirements:
        - host: lb_server

    master_server:
      type: tosca.nodes.indigo.Compute
      capabilities:
        endpoint:
          properties:
            dns_name: mesosserverpublic
            network_name: PUBLIC
        scalable:
          properties:
            count: 1

    mesos_slave_server:
      type: tosca.nodes.indigo.Compute

    lb_server:
      type: tosca.nodes.indigo.Compute
      capabilities:
        endpoint:
          properties:
            network_name: PUBLIC
        scalable:
          properties:
            count: 1

  outputs:
    mesos_lb_ips:
      value: { get_attribute: [ lb_server, public_address ] }
    mesos_master_ips:
      value: { get_attribute: [ master_server, public_address ] }
\end{lstlisting}
%\end{minipage}

\section{Conclusions and future work}
\label{sec:conclusions}

This paper has described the challenges of orchestrating computing resources
in heterogeneous clouds and the approach carried out in the INDIGO-DataCloud
project to overcome them, with discussion of real life use cases from
scientific communities.

The TOSCA open standard has been adopted for the description of the
application layouts. Different examples have been shown ranging from a simple
single-node application to an elastic Apache Mesos cluster. For this, an orchestration approach based
on prioritizing cloud sites with existing pre-configured Docker images is
employed, while being able to dynamically deploy the applications on cloud
sites  supporting only vanilla VMs or Docker images. By adopting a
configuration management solution based on Ansible roles to carry out both
the deployment of the application and the creation of the pre-configured Docker
images, a single consistent unified approach for application delivery is employed.

By using TOSCA to model the user's complex application architectures it is
possible to obtain repeatable and deterministic deployments. User's can port
their virtual infrastructures between providers transparenty obtaining the
same expected topology.

The time required to deploy a virtual infrastructure is strictly dominated by
the time required to provision the underlying computational resources and the
time to configure them. Therefore, provisioning from a Cloud site that already
supports the pre-configured Docker images requested by the user is considerably
much faster than having to boot up the Virtual Machines from another Cloud
site and perform the whole application installation and its dependencies.
The overhead introduced by the PaaS layer is negligible compared to the
time to deploy an infrastructure, since it just requires a reduced subset
of invocations among the different microservices.

As it can be seen from the use cases described in Section~\ref{sec:use cases},
several non-normative TOSCA node types were introduced in the context of INDIGO-DataCloud
project to support both different user applications and specific services to be used
within the deployed applications. Indeed, the extensibility of the TOSCA language and
the ability of the underlying TOSCA parser to process these new elements
facilitates the procedure of adopting TOSCA as the definition language to
perform the orchestration of complex infrastructures across multiple Clouds.

It is important to point out that a key contribution of the INDIGO-DataCloud
Orchestration system, with respect to other orchestration platforms, is the
implementation of hybrid orchestration to satisfy demands of dynamic or highly
changeable workloads, such as the virtual elastic cluster use case presented.

The approach described here is being used by several user communities that have been
engaged within the project \cite{davidoviceuropean,plociennik2016two,fiore2016distributed,owsiak2017running,aguilar2017hydrodynamics,fiore2017big,chen2016indigo,monna2017emso,kurkcuoglu2016science}. The developed solutions have also resulted in community and
upstream code contributions to major open source solutions like OpenStack and
OpenNebula.

%Also, the ability to perform hybrid deployments across multiple cloud
%sites specified in the TOSCA template as part of the requirements of the applications.
Future work includes supporting different complex application  architectures
used by scientific communities. For example, this work will include architectures
for Big Data processing in order to automatically provision virtual computing
clusters to process large volumes of data using existing frameworks such as Hadoop
and Spark). Regarding the tools that have been described, the OpenStack Heat Translator is
expected to evolve into a service that could be deployed along with the OpenStack
Orchestration (Heat) service, enabling the direct submission of TOSCA documents
to the endpoint that this service will provide.

\begin{acknowledgements}
    The authors want to acknowledge the support of the INDIGO-Datacloud (grant
    number 653549) project, funded by the European Commission's Horizon 2020
    Framework Programme.
\end{acknowledgements}


\bibliographystyle{spmpsci}
\bibliography{references}

\end{document}